\documentstyle[graphicx,amsmath,amssymb,fleqn]{mn}

\newif\ifAMStwofonts
\AMStwofontstrue
\DeclareMathAlphabet{\mathfrak}{U}{euf}{m}{n}
\SetMathAlphabet\mathfrak{bold}{U}{euf}{b}{n}
\newcommand{\bfr}{{\bmath r}}
\newcommand{\bfv}{{\bmath v}}

\newcommand{\micron}{{\mbox{\,$\mu$m}}}

\makeatletter
  \newcommand\figcaption{\def\@captype{figure}\caption}
  \newcommand\tabcaption{\def\@captype{table}\caption}
\makeatother

\title[Kinematics of dusty ellipticals II]%
{Kinematics of elliptical galaxies with a diffuse dust component -- II.\ Dust effects on kinematic modelling}

\author[Baes, Dejonghe \& De Rijcke]%
{Maarten Baes$^{1,2}$, Herwig Dejonghe$^1$ and Sven De Rijcke$^{1,2}$\\
$^1$Sterrenkundig Observatorium Universiteit Gent, Krijgslaan 281 S9, B-9000 Gent, Belgium \\
$^2$Research Assistent of the Fund for Scientific Research - Flanders (Belgium)
} 

\begin{document}

\maketitle
\begin{abstract}
Observations have demonstrated the presence of substantial amounts of
interstellar dust in elliptical galaxies, most of which is believed to
be distributed diffusely over the galaxy. Absorption by dust grains
has a major impact on the transfer of stellar radiation, and affects
the projection of each physical (and kinematic) quantity. In a
previous paper, we have investigated the effects of a diffusely
distributed dust component on the observed kinematics of spherical
galaxies. In this paper we investigate the effect of not taking dust
into account in dynamical modelling procedures. We use a set of
semi-analytical dusty galaxy models to create synthetic LOSVD data
sets, which we then model as if no dust were present.

We find some considerable differences between the best-fitting models
and the input models, and we find that these differences are dependent
on the orbital structure of the input galaxy. For radial and isotropic
models on the one hand, we find that the dynamical mass of the models
decreases nearly linearly with optical depth, with an amount of 5~per
cent per optical depth unit, whereas their orbital structure is hardly
affected. For tangential models on the other hand, the dynamical mass
decrease is smaller, but their orbital structure is affected~: their
distribution functions appears less tangentially anisotropic. For all
models the mass-to-light ratio will be underestimated, by a factor of
around 20~per cent per optical depth unit.

We discuss these results in the light of the limited effects of dust
extinction on the LOSVDs, as obtained in paper~I, and conclude that
the determination of the dynamical mass and the kinematic structure of
galaxies is not only determined by the observed kinematic quantities,
but is also critically dependent on the potential and hence the
observed light profile. We argue that dust, even in rather modest
amounts, should therefore be taken into account in kinematic modelling
procedures.
\end{abstract}



\section{Introduction}

It has become well-established that early-type galaxies contain a
considerable amount of interstellar dust. In the optical dust is
detected in ellipticals by its obscuration effects on the light
distribution, when it is present in the form of dust lanes and patches
(e.g.\ Ebneter \& Balick 1985, V\'eron-Cetty \& V\'eron 1988, van
Dokkum \& Franx 1995). In emission, dust is detected by the IRAS
satellite in the 60 and 100\micron\ wavebands (Knapp {\em et al.\
}1989, Roberts {\em et al.\ }1991). In a comparative analysis
Goudfrooij \& de Jong (1995, hereafter GdJ95) show that the dust
masses derived from FIR data are about a factor of ten higher than
those calculated from optical data. Using a more detailed dust mass
estimator which includes a temperature distribution for the dust
grains (Kwan \& Xie 1992), Merluzzi (1998) shows that GdJ95 still
underestimated the FIR dust masses with a factor up to six. Also
submillimeter observations (Fich \& Hodge 1991, 1993, Wiklind \&
Henkel 1995) and ISO data beyond 100\micron\ (Haas 1998), that may be
able to detect the very cold dust for which IRAS is insensitive
($T<25$\,K), suggest that the dust masses could be up to an order of
magnitude higher than observed from FIR observations alone.

This difference between the absorption and emission dust masses in
elliptical galaxies is called the dust mass discrepancy. It cannot be
solved by a more critical reconsideration of the IRAS data (Bregman
{\em et al.\ }1998), by corrections for the optical absorption in the
dust lanes (Merluzzi 1998) or by taking into account the dust recently
ejected from evolved stars (Tsai \& Mathews 1996). The interstellar
dust medium in ellipticals hence has to be composed of (at least) two
components: a lesser massive one which is optically visible in the
form of dust lanes and patches, and a more massive one which is
distributed over the galaxy and is hard to detect optically. The
spatial distribution of this component is still unclear. GdJ95 suggest
that it is distributed over the inner parts of the galaxy, which
supports the evaporation flow scenario. In this picture, most of the
gas and dust has an external origin: clouds of interstellar matter
have been accreted during interactions/merging with a gas-rich galaxy,
and they gradually evaporate in the hot X-ray emitting gas (Sparks,
Macchetto \& Golombek 1989, de Jong {\em et al.}  1990, Forbes 1991,
GdJ95). Tsai \& Mathews (1996) and Wise \& Silva (1997) on the other
hand suggest that dust is not confined to the inner few kpc only, but
also extends to larger radii. Such a dust distribution supports a
scenario where gas and dust have an internal origin: it can be
associated with red giant winds or formed in connection with star
formation in cooling flows (Fabian, Nulsen \& Canizares 1991; Knapp,
Gunn \& Wynn-Williams 1992; Hansen, J{\o}rgensen \&
N{\o}rgaard-Nielsen 1995). However, none of these mechanisms seems to
be able to explain the FIR observations satisfyingly, and a
combination of both mechanisms may be at work (Merluzzi 1998). As in
the first paper of this series (Baes \& Dejonghe 2000, hereafter
paper~I), we will assume that dust is distributed smoothly over the
entire galaxy.

Absorption by dust grains has a major impact on the transfer of
stellar radiation through the interstellar medium. It therefore
affects the projection, i.e.\ the integration along the line-of-sight
(hereafter LOS) of the intrinsic three-dimensional light
distribution. A number of studies have investigated the effects of a
diffuse dust distribution upon the photometry of ellipticals (Witt,
Thronson \& Capuano 1992, GdJ95, Silva \& Wise 1996, Wise \& Silva
1996; hereafter WS96). But dust does not only affect the projection of
the light distribution, it affects the projected kinematics too. This
is of particular importance in stellar dynamics, where the ultimate
purpose is the determination of the phase space distribution function
$F(\bfr,\bfv)$, hereafter DF, describing the entire dynamical
structure. This DF is usually constructed by fits to the LOSVDs or
their moments, and hence it is obviously important to examine to which
extent these are affected by dust obscuration.

In paper~I we constructed a set of semi-analytical models in order to
investigate the effects of dust absorption on the light profile, the
projected velocity dispersion profile and the LOSVDs. Concerning the
photometry, we find that diffuse dust has a strong impact~: a global
attenuation, strong extinction in the central regions, the formation
of radial color gradients, and an increasing apparent core radius. In
spite of the simplified character of our analysis (we don't include
scattering effects in our models), these results are in qualitative
correspondence with the conclusions of the more detailed photometric
studies mentioned before. The effect of dust on the projected
dispersion profile or the LOSVDs is of a totally different nature. The
effects are a redistribution along the LOS: dust makes the
contribution of the nearer parts more important, relative to the more
distant parts. Therefore it is no big surprise that, in a spherically
symmetric galaxy, the LOSVDs are quite insensitive to modest amounts
of dust. For example, for a modest optical depth ($\tau=2$) the effect
on the projected velocity dispersion is around 2~percent in the
central regions. However, for higher optical depths, these effects do
become considerable, which can be important on a local scale: e.g.\
asymmetries in the projected kinematics may be the result of a large
extinction in a compact region, such as in the dust lane elliptical
NGC\,5266 (M\"ollenhof \& Marenbach 1986). More details can be found
in paper~I.

The fact that both the photometry and the observed kinematics are
affected by dust obscuration will have consequences on the dynamical
modelling of galaxies. Now that we understand the way dust absorption
affects the {\em projected} kinematics, it is obvious that dust needs
to be accounted for in {\em deprojection} procedures, or more general,
in kinematic modelling procedures. It is the scope of this paper to
determine the effects of neglecting dust in the modelling of kinematic
data. In other words, if one tries to model dust-affected kinematic
profiles without taking dust into account, as is usually done, to what
extent are erroneous conclusions drawn about the kinematic structure
of the galaxy ? In order to answer this question we will create
synthetic dust-affected data sets, and model them as if no dust were
present.

In section~2 and section~3 we describe the construction of the data
sets and the modelling procedure. The results of the modelling are
presented in section~4, and in section~5 we investigate whether these
depend critically on the choice of the dust geometry. A discussion of
the results is given in section~6.


\section{The data sets}


\subsection{The model}
\label{themodel.par}

To create our data sets, we construct a set of two-component galaxy
models, consisting of a stellar and a dust component. Both components
are spherically symmetric.

For the stellar component, we use a Plummer model (Dejonghe 1987),
which is described by the potential-density pair
\begin{subequations}
\begin{gather}
	\psi(r) = \frac{GM_0}{c}\,
		  \left(1+\frac{r^2}{c^2}\right)^{-\frac{1}{2}}
	\label{plummpair1} \\
	\rho(r) = \frac{3}{4\pi}\,\frac{M_0}{c^3}\,
		  \left(1+\frac{r^2}{c^2}\right)^{-\frac{5}{2}},
	\label{plummpair2}
\end{gather}
\end{subequations}
with $c=5$\,kpc the so-called core radius and
$M_0=5\times10^{10}$\,M$_\odot$ the total mass. Furthermore we assume
a constant mass-to-light ratio $\Upsilon(r) = 4\,\Upsilon_\odot$, such
that the three-dimensional light distribution is given by
$\ell(r)=\tfrac{1}{4\,\Upsilon_\odot}\,\rho(r)$.

With the technique described by Dejonghe (1986) one finds a family of
two-integral DFs $F_q(E,L)$ that self-consistently generate the
Plummer potential-density pair~(\ref{plummpair1}b). This family has
two interesting properties, which justify its choice as a generic
model for the class of elliptical galaxies (although its projected
density profile does not fit real elliptical galaxies). First, the
models are completely analytical, i.e.\ the DF and all the (projected)
moments can be calculated analytically. And second, the family depends
on one single parameter $q$ which can be varied continuously in order
to obtain tangential ($q<0$), isotropic ($q=0$) or radial ($0<q<2$)
models. We consider a set of different orbital structures,
characterized by the parameters $q=-6$, $-2$, $0$ and $1$. In the
sequel of this paper we will refer to them as very tangential,
tangential, isotropic and radial models respectively.

As in paper~I, we only incorporate the effects of dust absorption and
neglect scattering effects. Then, the dust component is completely
determined by the opacity function $\kappa(r)$, for which we use a
modified Hubble profile
\begin{equation}
	\kappa(r)
	=
	\frac{\tau}{2\,c}\,
	\left(1+\frac{r^2}{c^2}\right)^{-\frac{3}{2}}.
\label{king_para}
\end{equation}
The normalization is such that $\tau$ equals the total optical depth,
defined as the projection of the opacity along the entire central
LOS\footnote[1]{Throughout paper~I and this paper we use this
definition of $\tau$, whereas e.g.\ WS96 defined $\tau$ as the
integral of the opacity from the centre of the galaxy to the edge,
half our value.}
\begin{equation}
	\tau 
	= 
	\int_{\text{central LOS}} \hspace{-1em}\kappa(r)\,{\rm d}s
	\,=\,
	2\int_0^{+\infty}\kappa(r)\,{\rm d}r.
\label{normalization}
\end{equation}
As WS96 and paper~I, we take the same core radius for the dust and the
stars; the choice of the opacity function is critically investigated
in chapter~5. We only consider optical depths ranging from $\tau=0$ to
$\tau=3$ in our calculations, as high values for the optical depth
associated with a diffuse dust component seem not to be in accordance
with photometric studies (GdJ95, WS96).

\subsection{The projections}
\label{projection.par}


\begin{table}
\centering
\caption{Some parameters of the dusty Plummer models as a function of
the optical depth $\tau$. The second column gives the observed core
radius, the third and the fourth column give the observed luminosity
and the total extinction, and column six and seven list the projected
light density and projected dispersion for the central LOS. This
dispersion is tabulated for the isotropic model, one finds the central
dispersion of the very tangential, tangential and radial models by
multiplying this with the factors $\sqrt{2}/2$, $\sqrt{3}/2$ and
$\sqrt{6/5}$ respectively.}
\begin{tabular}{cccccc}
\label{param.tbl}
$\tau$ & $c_{\text{obs}}$ & $L_{\text{obs}}$ & $A$ & $\ell_{p,0}$ & $\sigma_{p,0}$ \\ 
& (kpc) & (10$^9$\,L$_\odot$) & (mag) & (L$_\odot$/pc$^2$) & (km/s) \\ \\
0.0 & 5.00 & 12.50 & 0.00 & 159.2 & 178.0 \\
0.5 & 5.33 & 11.08 & 0.13 & 124.7 & 177.8 \\
1.0 & 5.65 &  9.91 & 0.25 &  99.0 & 177.6 \\
1.5 & 6.02 &  8.93 & 0.37 &  79.5 & 177.1 \\
2.0 & 6.35 &  8.11 & 0.47 &  64.6 & 176.5 \\
2.5 & 6.74 &  7.41 & 0.57 &  53.1 & 175.7 \\
3.0 & 7.15 &  6.81 & 0.66 &  44.2 & 174.9
\end{tabular}
\end{table}


For each model we create a set of so-called {\em dusty} projected
kinematic data. A dusty projected quantity $\mu_p(x,\bfv)$ differs
from a normal projected quantity as it is a weighted integral along
the LOS $x$ of a three-dimensional spherically symmetrical quantity
$\mu(r,\bfv)$.  We assume that the galaxy is located at a distance
which is significantly larger than its size. Then the errors made by
assuming parallel projection, which are of the order $(c/D)^2$ with
$D$ the distance to the galaxy, are negligible (see Paper~I,
section~2). For a dusty galaxy with opacity function $\kappa(r)$ the
appropriate formula reads
\begin{subequations}
\begin{equation}
	\mu_p(x,\bfv)
	=
	2\int_x^{+\infty}\!{\cal K}(x,r)\,
	\frac{\mu(r,\bfv)\,r\,{\rm d}r}{\sqrt{r^2-x^2}},
\label{dusty_projection}
\end{equation}
where ${\cal K}(x,r)$ is a weight function defined as 
\begin{gather}
	{\cal K}(x,r) 
	=
	\exp\left(-\int_x^{+\infty}
	\frac{\kappa(r)\,r\,{\rm d}r}{\sqrt{r^2-x^2}}\right) 
	\nonumber \\
	\qquad\qquad\qquad\qquad \times \quad
	\cosh\left(\int_x^r
	\frac{\kappa(r')\,r'\,{\rm d}r'}{\sqrt{{r'}^2-x^2}}\right).
\label{weight_function}
\end{gather}
\end{subequations}
Details can be found in section~2 of paper~I. We substituted the
analytical expressions for the moments of the DF, together with the
opacity function~(\ref{king_para}), in
expression~(\ref{dusty_projection}) to obtain dusty projected
profiles, such as the projected light density $\ell_p(x)$ and the
projected velocity dispersion $\sigma_p(x)$. Since in particular the
projected light density will depend on the optical depth, we will have
to be careful in our terminology when calculating quantities such as
the luminosity and the core radius. We need to discriminate between
{\em true} and {\em observed} quantities, which are respectively
derived from the spatial and the projected distribution. For example,
the true luminosity of the galaxy is calculated by integrating the
light density over space
\begin{equation}
	L = 4\pi 
	\int_0^{+\infty} \ell(r)\,r^2\,{\rm d}r, 
\end{equation}
which is of course independent of the optical depth and equals
$M_0/\Upsilon = 1.25\times10^{10}$~L$_\odot$. The observed luminosity
$L_{\text{obs}}$ is calculated by integrating the projected light
density over the plane of the sky,
\begin{equation}
	L_{\text{obs}} = 2\pi
	\int_0^{+\infty} \ell_p(x)\,x\,{\rm d}x,
\end{equation}
and is a function of $\tau$. Analogously, all our Plummer models have
a (true) core radius $c=5$\,kpc, whereas their observed core radius
$c_{\text{obs}}$ will depend on the optical depth (see
section~{\ref{potential.par}}). In table~{\ref{param.tbl}} we tabulate
some of the parameters of our dusty Plummer models.

\subsection{The data sets}

Each data set consists of the projected light density data $\ell_p(x)$
and projected dispersion data $\sigma_p(x)$, which are taken at
$x=0$\,kpc, 0.5\,kpc,\ldots,10\,kpc. However, since dispersion
profiles depend on both the orbital structure and the mass
distribution, they do not sufficiently constrain the dynamical
structure of the galaxy. This mass-anisotropy degeneracy can be broken
by including the higher order Gauss-Hermite moments in the fitting
routine (van der Marel \& Franx 1993, Gerhard 1993). Since we work
with simulated data, we are able to include LOSVD data points directly
in the modelling procedure. We assume that these data are noiseless,
such that our data set is perfect, i.e.\ it contains all the kinematic
information that could be available from perfect
observations. Altogether each data set consists of 252 data points.


\section{The modelling procedure}


\subsection{Determination of the potential}
\label{potential.par}


\begin{figure}
\centering 
\includegraphics[clip,bb=182 518 412 685]{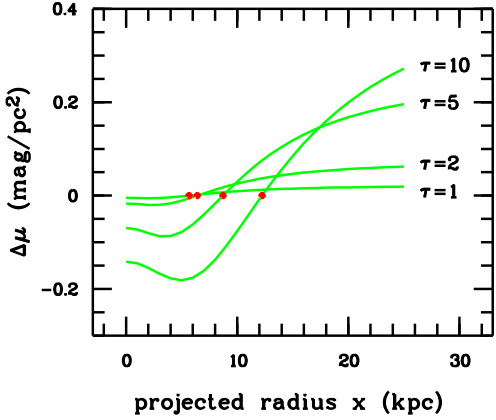}
\caption{The difference $\Delta\mu(x)$ in surface brightness between
our dusty projected density profiles and the best fitting Plummer
light profiles, for the modest optical depths $\tau=1$ and $\tau=2$,
and also for the higher values $\tau=5$ and $\tau=10$. It is clear
that for modest optical depths a Plummer profile provides a satisfying
fit to the dust-affected profiles, such that we can safely assume a
Plummer potential in our modelling procedure. The dots in the figure
indicate the observed core radii $c_{\text{obs}}$ of the Plummer light
profiles, where by construction $\ell_p(c_{\text{obs}}) =
\ell_p^{\text{Pl}}(c_{\text{obs}})$. }
\label{plprof.ps}
\end{figure}


The first step in the modelling of kinematic data is the determination
of the potential. Although we have a perfect data set at our disposal,
we are not able to constrain the potential completely without any
assumptions, not even in the case of spherical symmetry (Dejonghe \&
Merritt 1992). In this first approach we assume that it makes sense to
consider a constant mass-to-light ratio, such that the spatial
dependence of the potential can be derived from the projected light
density $\ell_p(x)$. It can be calculated numerically for our set of
models, but we prefer to work with a potential whose functional form
is explicitly known. We find that, for the modest optical depths we
are considering, the dust-affected light profiles of our models can
still be satisfyingly described by Plummer light profiles
\begin{equation}
	\ell_p^{\text{Pl}}(x) 
	=
	\ell_{p,0}\,
	\left(1+\frac{x^2}{c_{\text{obs}}^2}\right)^{-2},
\label{plprof}
\end{equation}
with $c_{\text{obs}}$ dependent on the optical depth. In
figure~{\ref{plprof.ps}} we plot the difference
\begin{equation}
	\Delta\mu(x) 
	=
	-2.5\log\left[\ell_p(x)/\ell_p^{\text{Pl}}(x)\right].
\end{equation}
in surface brightness between the dusty projected light density
profiles $\ell_p(x)$ and the best fitting Plummer light profiles
$\ell_p^{\text{Pl}}(x)$. Even at very large projected radii we find
that $\Delta\mu$ never becomes larger than 0.05 for $\tau=2$, such
that we can say that the Plummer character of the galaxy is preserved
for modest values of $\tau$. We will therefore assume a Plummer
potential
\begin{equation}
	\psi(r) 
	= 
	\frac{GM}{c_{\text{obs}}}\,
	\left(1+\frac{r^2}{c_{\text{obs}}^2}\right)^{-\frac{1}{2}}
\label{Plpot}
\end{equation}
for our models, with $c_{\text{obs}}$ determined from the best fit to
the $\ell_p(x)$ data, and the mass $M$ still featuring as a free
parameter. The values for $c_{\text{obs}}$ are tabulated in the second
column of table~1.

\subsection{Determination of the DF}

With a fixed potential there exists one and only one two-integral DF
$F(E,L^2)$ that fits the kinematic data (Dejonghe \& Merritt 1992). It
is always possible to write this DF as a infinite sum of components
\begin{equation}
	F(E,L^2) = \sum_{i=1}^{\infty} c_i F^i(E,L^2)
\end{equation}
where $c_i$ are the coefficients and the components $F^i(E,L^2)$ form
a complete set of simple dynamical models. For any observed kinematic
data point $\mu_n(x,\bfv)$ the same expansion is valid,
\begin{equation}
	\mu_n = \sum_{i=1}^{\infty} c_i \mu_n^i,
\end{equation}
since these moments depend linearly on the DF. Practically one can
only consider a finite number $N$ of components,
\begin{equation}
	F(E,L^2) \approx \sum_{i=1}^{N} c_i F^i(E,L^2).
\end{equation}
The best fitting coefficients can then be found by minimizing a
$\chi^2$-like variable
\begin{equation} 
	\chi^2 = \sum_{n}
	\left[ w_n
	\left( 1-\frac{\sum_{i=1}^{N} c_i \mu_n^i}{\mu_n} \right)
	\right]^2
\end{equation}
where the sum contains all data points, and $w_n$ is the weight
accorded to the $n$th data point. Since we assume to obtain perfect,
noiseless synthetic data, we can use these weights to (arbitrarily)
set the relative importance of each data point in the global
$\chi^2$. For the projected density points we take $w_n=1$, for the
projected dispersion $w_n=\mbox{$\frac{1}{3}$}$ and for the LOSVD data
the weight varies from $w_n=\mbox{$\frac{1}{3}$}$ at the centre to
$w_n=\mbox{$\frac{1}{10}$}$ in the outer parts. This $\chi^2$ is
quadratic in the coefficients and has to be minimized under the linear
constraint that the DF has to be positive over a grid $(E_j,L^2_k)$ in
phase space,
\begin{equation}
	\sum_{i=1}^{N} c_i F^i(E_j,L^2_k) \ge 0 \quad 
	\mbox{for all $j$ and $k$,}
\end{equation}
which amounts to a typical Quadratic Programming problem. For details
we refer to Dejonghe (1989).

We choose our components from a library of Fricke models. These are
simple dynamical models that are defined by the augmented
density\footnote[1]{The augmented density is the density written as a
function of the radius $r$ and the potential $\psi$. It is a
fundamental quantity in a technique to construct 2-integral
distribution functions (Dejonghe 1986).},
\begin{equation}
	\rho(r,\psi) 
	= 
	\psi^a \left(\frac{r}{s_*}\right)^{2b}. 
	\label{augm}
\end{equation}
where $s_*$ is a scale factor and $a$ and $b$ are real numbers that
satisfy the condition $a-2b>3$ to keep the total mass
finite. The anisotropy $\beta(r)$, generally defined as
\begin{equation}
\label{defaniso}
	\beta(r) 
	=
	1- \frac{\sigma^2_\varphi}{\sigma^2_r},
\end{equation}
will be constant for these models, since the $r$-dependence is a power
law (Dejonghe 1986, section~1.5.1). We find immediately $\beta(r)=-b$,
hence the condition $b>-1$ is required. The two-integral DF
corresponding to~(\ref{augm}) is a simple power-law of $E$ and $L^2$
\begin{equation}
	F(E,L^2) \propto E^{a-b-\frac{3}{2}}L^{2b}.
\end{equation}
For tangentially anisotropic components the central density vanishes,
for isotropic ones it is finite and non-zero and radially anisotropic
components have a central density cusp. The advantage of this family
lies within the fact that most of the kinematics can be calculated
analytically for positive, integer values of $b$ (De Rijcke \&
Dejonghe 1998).

\subsection{Determination of the dynamical mass}
\label{determass.par}

The only unknown in our model now is the mass $M$, which still acts as
a free parameter. For its determination we run our models for a number
of possible values, and determine the best fitting DF and the
according $\chi^2$ parameter for each value. The best fitting mass of
the model is then determined as the minimum in $\chi^2(M)$.

\subsection{Practical application}


\begin{figure}
\centering 
\includegraphics[clip,bb=182 518 412 685]{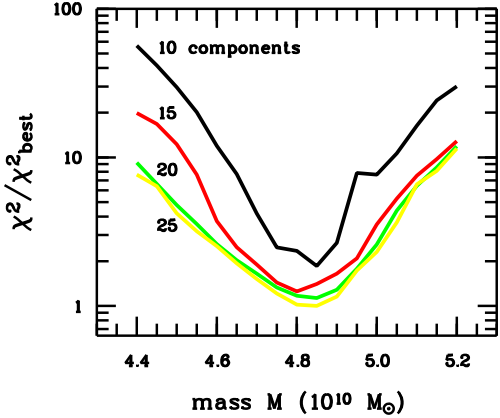}
\caption{Plot showing $\chi^2$ as a function of the mass $M$, for a
tangential input model with $\tau=1$. The $\chi^2$ values have no
absolute meaning. The dynamical mass is determined at the minimum of
the plot, being $4.85\times10^{10}$\,M$_\odot$. The plot is shown for
models containing 10, 15, 20 and 25 components. Clearly 20~components
are sufficient to determine the dynamical mass accurately.}
\label{detmass.ps}
\end{figure}


\begin{figure*}
\centering 
\includegraphics[clip,bb=65 55 540 625,width=175mm]{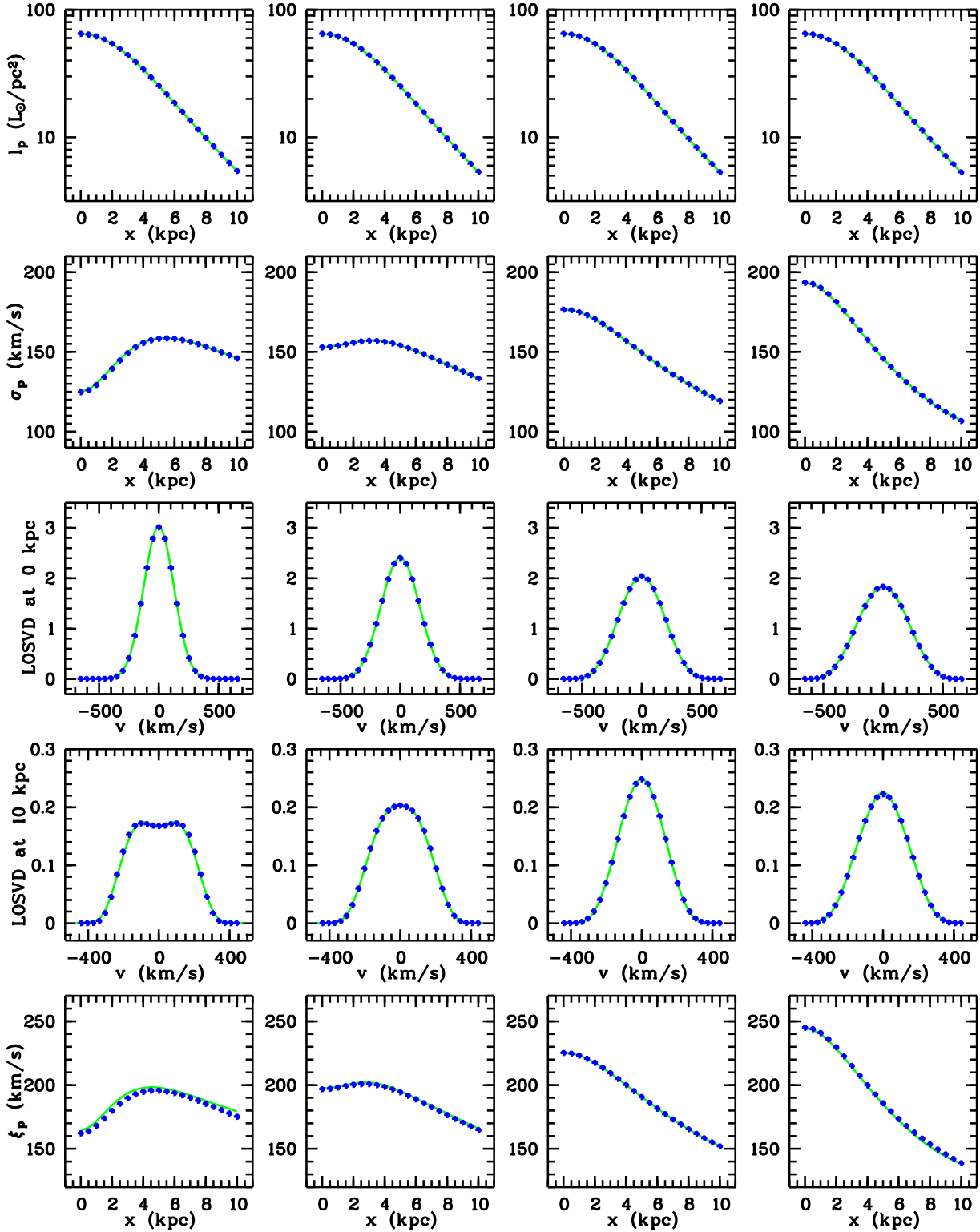}
\caption{Results of the fitting procedure for the very tangential
(left), tangential (middle left), isotropic (middle right) and radial
(right) models, with optical depth $\tau=2$. Shown are the projected
light density profile $\ell_p(x)$ (in arbitrary units), the projected
dispersion profile $\sigma_p(x)$ and the LOSVDs for $x=0$\,kpc and
$x=10$\,kpc. The dots are the data points, the solid lines represent
the fit. The bottom row shows the projected fourth moment $\xi_p(x)$
of the LOSVDs, which is not included in the fit and can be used as a
check on the results. }
\label{fits.ps}
\end{figure*}


Practically we construct models for a sufficiently large set of
masses, with a step $\Delta M = 5\times10^8$\,M$_\odot$, 1 per cent of
the input mass. For each mass we construct a DF using a library of
30~isotropic and tangential Fricke components. For computational
reasons we do not include any radial components; radial models can be
constructed by linear combinations of isotropic and tangential
components, in which the latter have a negative weight (allowed as
long as the DF remains positive over phase space). We typically used
20~components: adding more components does not significantly affect
the results (neither the DF nor the mass) anymore. This is illustrated
in figure~{\ref{detmass.ps}}, where we plot the $\chi^2$ values of the
various fits in function of the number of components and the value of
the total mass.

A check on the fitting procedure is done by using $\tau=0$ data, i.e.\
data from non-dusty Plummer models. In order to make the test robust,
we use the same strategy as De Rijcke \& Dejonghe (1998): for the
Plummer models that can be fit exactly in terms of Fricke components
($q=-6$, $-2$ and 0), we remove these components from the
library. Still, both the DF and the mass of the input models can be
successfully reproduced.

In figure~{\ref{fits.ps}} we plot some results of the fitting
procedure for the four $\tau=2$ models. Shown are the projected light
density $\ell_p(x)$, the projected dispersion $\sigma_p(x)$ and the
LOSVDs at $x=0$~kpc and $x=10$~kpc, the innermost and outermost LOSVD
in our dataset. The quality of the fit cannot be deduced from the
$\chi^2$ values, since these have no statistical meaning.  At bottom
row in figure~{\ref{fits.ps}} we show the projected fourth moment
$\xi_p(x)$ of the DF, which is not included in the modelling
procedure. It can be used to check the quality of the fit, which is
very satisfactory in all cases.


\section{Results}


In this section we describe the results of our fitting procedure and
the kinematic properties of the models. For the sake of clarity we
first explicitly define some terms.

The {\em input models} are the models that are described in
section~{\ref{themodel.par}}, i.e.\ Plummer galaxy models containing a
dust component, whereas the {\em fitted models} are the models that
come out of the modelling procedure, and which, by construction,
contain no dust. Since each couple of parameters $(q,\tau)$
corresponds to one input model, and hence one data set and one fitted
model, we will call the combination of input and fitted models
corresponding to a given couple of parameters simply a {\em model}.

As the fitted models are constructed such that their projected
kinematics match these of the input models, we can talk about {\em
the} projected kinematics of a model. The same applies to the light
profile or quantities derived from it, such as the observed luminosity
$L_{\text{obs}}$ (see section~{\ref{projection.par}}). On the
contrary, when we describe spatial kinematic quantities (such as the
anisotropy $\beta(r)$) or integrals thereof (such as the mass-to-light
ratio $\Upsilon$), we need to distinguish between the ones
corresponding to the input and fitted models, which a priori have no
reason to be equal. With an {\em apparent quantity} we mean a quantity
that corresponds to the fitted model, e.g.\ the quantity that results
from the modelling procedure. With an {\em intrinsic quantity} we mean
the quantity that corresponds to the input model, and hence is always
independent of the optical depth. For example, all the models have the
same intrinsic mass $M_0$, whereas the apparent mass of the models is
determined as outlined in section~{\ref{determass.par}}, and will be
different for each model. Obviously, for optical depth $\tau=0$ the
apparent and intrinsic values are equal.

In this section we will compare the apparent and the intrinsic
kinematic quantities of our models, as a function of the parameters
$\tau$ and $q$.

\subsection{The mass and the mass-to-light ratio}


\begin{figure*}
\centering 
\includegraphics[clip,bb=40 488 554 685]{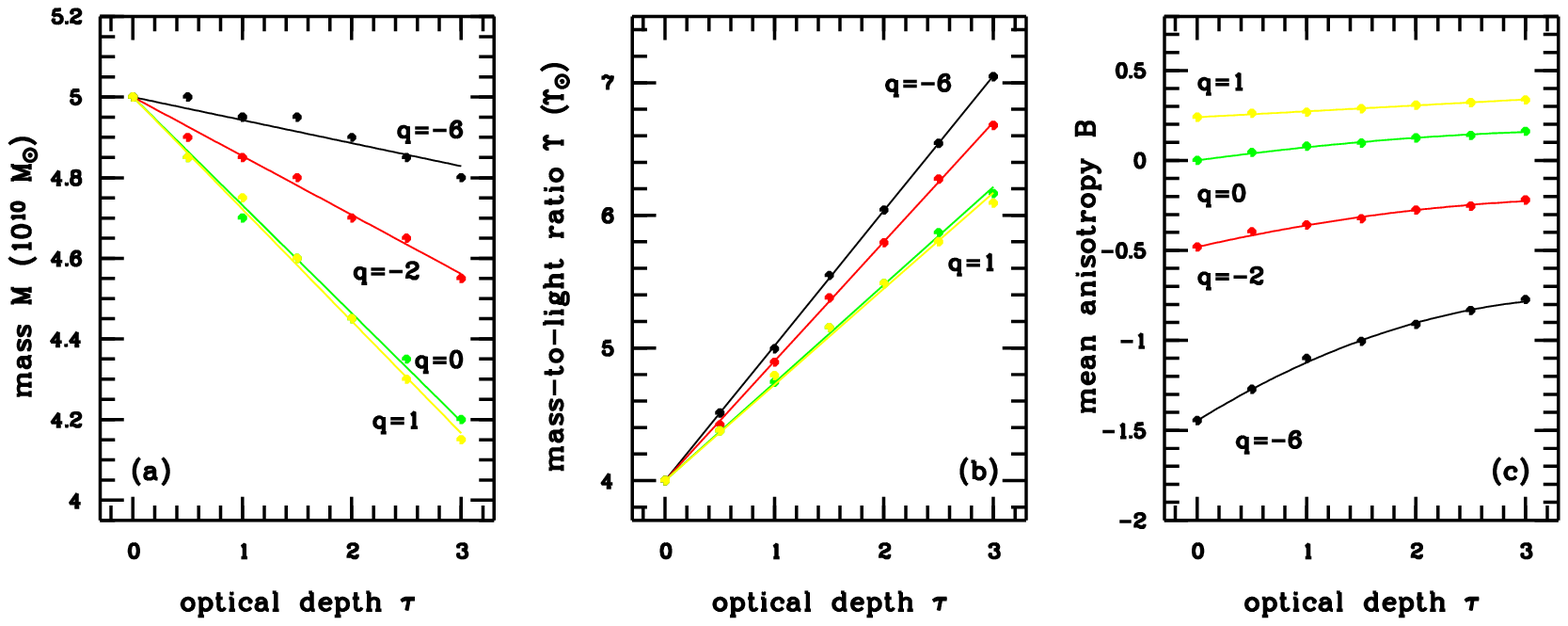}
\caption{Three plots, showing the apparent mass $M$, the apparent
mass-to-light ratio $\Upsilon$ and the apparent mean anisotropy
${\mathcal{B}}$ of our models, as a function of the optical depth
$\tau$. They are shown for different orbital structures, ranging
from very tangential (black) to radial (light grey). The dots
represent the results of our fits, the solid lines are least-square
fits to these points, either linear (for the mass and the
mass-to-light ratio) or quadratic (for the mean anisotropy). The
corresponding coefficients are tabulated in
table~{\ref{deptau.tbl}}. }
\label{deptau.ps}
\end{figure*}


\begin{table}
\centering
\caption{The parameters $a_M$, $a_\Upsilon$, $a_{\mathcal{B}}$ and
$b_{\mathcal{B}}$ from the least-squares fits to $\Delta M$,
$\Delta\Upsilon$ and $\Delta{\mathcal{B}}$ (see text). For $a_M$ and
$a_\Upsilon$ we give both the absolute and relative values.}
\label{deptau.tbl}
\begin{tabular}{rcccc}
$q$ & $a_M$ & $a_\Upsilon$ & $a_{\mathcal{B}}$ & $b_{\mathcal{B}}$ \\ 
& (10$^9$ M$_\odot$) & ($\Upsilon_\odot$) & & \\ \\
-6 & 0.57 (1.1\%) & 1.02 (25\%) & 0.65 & -0.09 \\
-2 & 1.46 (3.0\%) & 0.90 (23\%) & 0.24 & -0.03 \\
 0 & 2.68 (5.4\%) & 0.74 (19\%) & 0.12 & -0.02 \\
 1 & 2.78 (5.6\%) & 0.72 (18\%) & 0.02 & -0.00 
\end{tabular}
\end{table}


In figure~{\ref{deptau.ps}a} we show the apparent dynamical mass of
the models as a function of the optical depth. The global effect of
the dust is clear~: the mass decreases nearly linearly with the
optical depth. The slope of this correlation however is strongly
dependent on the orbital structure. For the very tangential model the
apparent mass is quite insensitive to the presence of dust, and the
impact of the dust extinction becomes gradually stronger as we move to
the tangential, the isotropic and the radial model. We fitted straight
lines $\Delta M = a_M\,\tau$ through the data to obtain characteristic
values for the mass decrease in function of the optical depth. These
values are given in the second column of table~{\ref{deptau.tbl}}.

The apparent luminosity of our models is calculated by integrating the
apparent light density $\ell(r)$ over space, or -- since the fitted
models contain no dust -- by integrating $\ell_p(x)$ over the plane of
the sky. It thus equals the observed luminosity of the dusty Plummer
models, which are tabulated in the third column of
table~{\ref{param.tbl}}.

Combining these with the apparent masses we can calculate the apparent
mass-to-light ratio $\Upsilon$ of our models. The apparent luminosity
decreases stronger than the apparent mass with increasing optical
depth, such that $\Upsilon$ is an increasing function of $\tau$. The
dependence of $\Upsilon$ on the orbital structure is only determined
by the apparent mass, as the apparent luminosities are independent of
the orbital mode. As a consequence, the dependence on $q$ is now
opposite: the mass-to-light ratio is most dramatically affected for
tangential models, and the effects are smaller for radial and
isotropic ones. The results are shown in
figure~{\ref{deptau.ps}b}. Again, the dependence on the optical depth
is nearly linear and straight lines $\Delta\Upsilon =
a_\Upsilon\,\tau$ are fitted through the data points to obtain
characteristic values. These are tabulated in the third column of
table~{\ref{deptau.tbl}}.

\subsection{The distribution function}


\begin{figure*}
\centering 
\includegraphics[clip,bb=40 433 564 718,width=175mm]{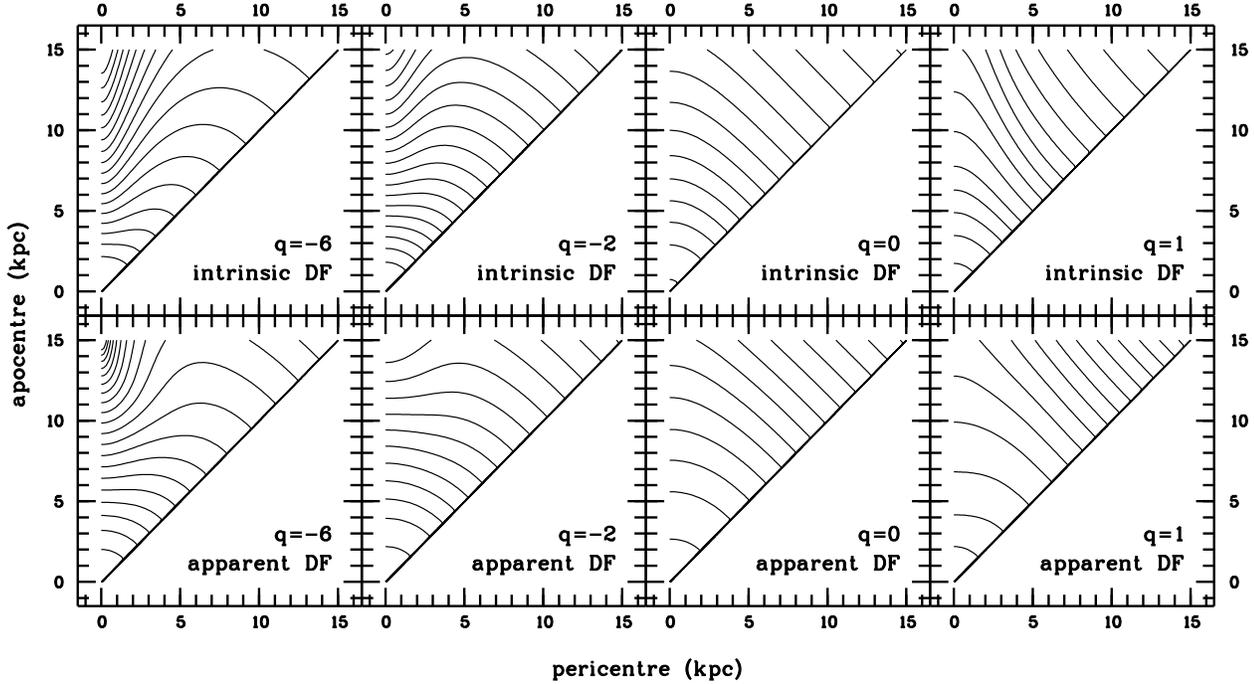}
\caption{Isoprobability contour plots in turning point space for the
$\tau=2$ models. The upper and lower panels represent the intrinsic
and the apparent DFs respectively. From left to right we have the very
tangential, tangential, isotropic and radial models, as in
figure~{\ref{fits.ps}}. }
\label{turnpoint.ps}
\end{figure*}


The eight panels in figure~{\ref{turnpoint.ps}} represent the
isoprobability contours of eight DFs, corresponding to the four
$\tau=2$ models. In the upper row we plot the intrinsic DFs, whereas
the lower panels represent the apparent DFs. From left to right we
have, as in the previous plots, the very tangential, tangential,
isotropic and radial models.

The contour plots are shown in turning point space, such that the DFs
can easily interpreted in terms of orbits. Let us first concentrate on
the four upper panels. In the innermost regions of the galaxies, the
shape of the DF is comparable -- indeed all Plummer models are fairly
isotropic in their centers. From a few kpc on however, we can clearly
see how the isoprobability contours reflect the orbital structure of
the model they represent. Tangential models prefer nearly circular
orbits, with a small difference between apocentre and pericentre, and
their contours will tend to lie alongside the diagonal axis. On the
other hand, radial models prefer elongated orbits, with a large
difference between apocentre and pericentre, such that their contours
will tend to be more vertical. The slope of the isoprobability
contours is thus indicative for the orbital structure of the model.

Let us now compare the intrinsic and apparent DFs. Concerning the
central regions we see that the same structure is preserved for all
orbital modes. Outside this region however, there are differences,
most clearly visible for the very tangential and tangential models~:
the contours are lying somewhat more horizontally, indicating that
elongated orbits are relatively more favored. Dust obscuration thus
seems to make these galaxies appear less tangential outside the
innermost regions. Whether a similar trend accounts for the isotropic
and radial models too, is less obvious from the DF contour
plots. Therefore we will study the anisotropy of our models in detail.

\subsection{The anisotropy}


\begin{figure*}
\centering 
\includegraphics[clip,bb=65 377 540 533,width=175mm]{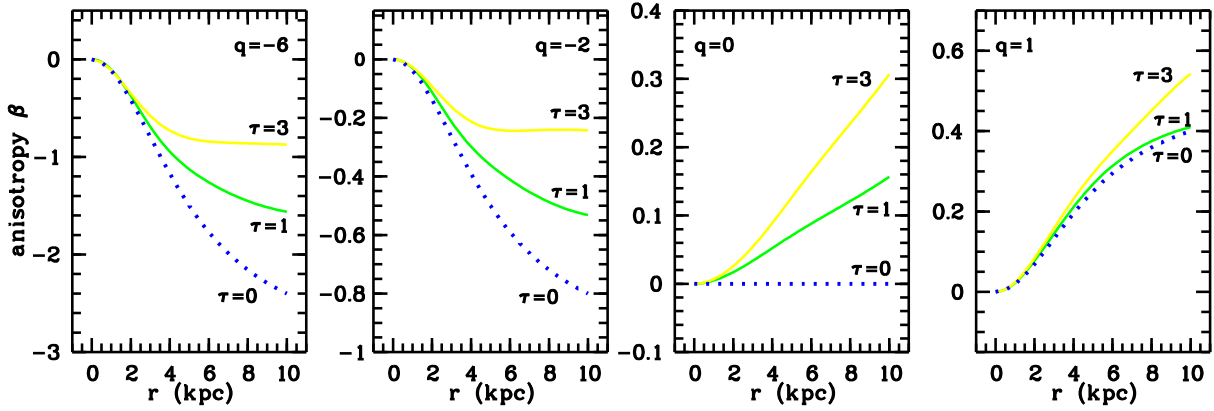}
\caption{The orbital structure of input and output models, as
characterized by the anisotropy $\beta(r)$. From left to right we have
a panel with the very tangential, tangential, isotropic and radial
models, as in figure~{\ref{fits.ps}}. The dotted lines show the
intrinsic orbital structure of the input galaxies, the solid lines
show the orbital structure of the $\tau=1$ and $\tau=3$ models. }
\label{orbstr.ps}
\end{figure*}


The intrinsic anisotropies of our models, as defined by
equation~(\ref{defaniso}), can be written as
\begin{equation}
	\beta(r) 
	= 
	\frac{q}{2}\,
	\frac{r^2}{r^2+c^2},
\label{anis}
\end{equation}
where $c$ represents the (true) core radius. All models are hence
intrinsically isotropic in the central regions, at least if isotropy
is defined only from the second order moments, and show their true
orbital behavior at larger radii.

The dependence of the apparent anisotropy on the optical depth is
shown, for the different orbital modes, in
figure~{\ref{orbstr.ps}}. Shown are apparent (solid lines) and the
intrinsic (dotted lines) anisotropies of the $\tau=1$ and $\tau=3$
models as a function of the spatial radius. For the very tangential
and tangential models, we see that the effect, as determined from the
DF plots, is confirmed: as for the intrinsic orbital structure, all
models are isotropic in their inner regions, and they are less
tangential at larger radii. Rather logically, this effect increases
with increasing optical depth. Looking at the third and fourth panel
we see that also the apparent anisotropy of the isotropic and radial
models increases outside a few kpc. All models thus seem to be subject
to a ``radialization'', i.e.\ dust obscuration tends to make galaxies
appear more radially anisotropic outside the central few kpc, even if
they are already intrinsically radial.

In order to quantify the strength of the radialization in function of
the input parameters $q$ and $\tau$, it is useful to consider one
single anisotropy parameter. We define a mean anisotropy
${\mathcal{B}}$ as
\begin{equation}
	{\mathcal{B}}
	=
	\frac{\int \beta(r)\,\rho(r)\,r^2\,{\rm d}r}
	{\int \rho(r)\,r^2\,{\rm d}r}
\label{meanbeta}
\end{equation}
with the integral covering the region of our fits (between 0 and
10~kpc). The intrinsic mean anisotropy for our models is off course
proportional to the parameter $q$; substituting~(\ref{anis})
and~(\ref{plummpair2}) in~(\ref{meanbeta}) we find ${\mathcal{B}} =
\tfrac{6}{25}\,q$. 

In figure~{\ref{deptau.ps}c} we plot the apparent mean anisotropy
${\mathcal{B}}$ for our models as a function of the optical depth. One
can clearly see that ${\mathcal{B}}$ increases for increasing optical
depth, and that the radialization is more dramatic the more tangential
the input model. The curves in the figure are quadratic fits
$\Delta{\mathcal{B}} = a_{\mathcal{B}}\,\tau +
b_{\mathcal{B}}\,\tau^2$ to the data points, and the coefficients are
tabulated in the last two columns of table~{\ref{deptau.tbl}}.


\section{Dependence on the dust model}


\begin{figure}
\centering 
\includegraphics[clip,bb=180 487 414 652]{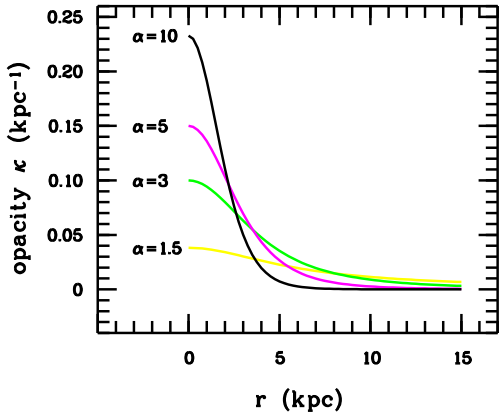}
\caption{The opacity function $\kappa(r)$, corresponding to 
equation~(\ref{kappa}) for different dust exponents $\alpha$. It is
shown for $\tau=1$, for other values of $\tau$ it scales linearly. For
large values of $\alpha$, the dust is concentrated in the central
regions, for large values the distribution is extended. }
\label{plotkappa.ps}
\end{figure}


The calculations in the previous chapters are based on a dust model
whose spatial dependence is given by equation~(\ref{king_para}). But,
as discussed in the introduction, very little is known about the
spatial distribution of the dust in elliptical galaxies. In this
chapter we will investigate whether the results so far obtained change
dramatically if the relative distribution of dust and stars
varies. Therefore we consider, as in paper~I, a more general opacity
function,
\begin{equation}
	\kappa(r)
	=
	\frac{1}{\sqrt{\pi}}\,\tau\,
	\frac{\Gamma\left(\frac{\alpha}{2}\right)}
	{\Gamma\left(\frac{\alpha-1}{2}\right)}\,
	\frac{1}{c}\,
	\left(1+\frac{r^2}{c^2}\right)^{-\frac{\alpha}{2}}
\label{kappa}
\end{equation}
which also satisfies the normalization
condition~(\ref{normalization}), and which reduces
to~(\ref{king_para}) if $\alpha=3$. The extra parameter in this family
of opacity functions, the dust exponent $\alpha$, sets the spatial
distribution of the dust. In figure~{\ref{plotkappa.ps}} we plot the
opacity function for different values of $\alpha$.

Small values of $\alpha$ correspond to spatially extended dust
distributions. The range of $\alpha$ is restricted to $\alpha>1$.
When $\alpha$ approaches this value, the dust is more or less equally
distributed along the line-of-sight, and as $c\ll D$, relatively very
little dust resides in the central regions of the galaxy. In the limit
$\alpha\rightarrow1$ the opacity function is such that the dust
effectively forms an obscuring medium between the galaxy and the
observer (see paper~I), analogous to extinction of starlight due to
interstellar dust in the Galaxy. This geometry is generally known as
the overlying screen approximation, and it is the geometry which has,
for a fixed optical depth, the largest impact on the projection of
starlight. For many years the extinction in spiral galaxies was
described using the cosecans law (Holmberg 1975), which implicitly
assumes this geometrical distribution of stars and dust. Nowadays
however, the extinction in these systems has been described using more
detailed dust-stars geometries, and the overlying screen model is
generally considered to be unsatisfying (Bruzual, Magris \& Calvet
1988; Disney, Davies \& Phillips 1989; Witt, Thronson \& Capuano
1992).

On the other hand, larger values of $\alpha$ correspond to centrally
concentrated dust. For $\alpha=5$ dust and stars have the same
geometry, and if $\alpha$ becomes very large the extinction is
confined to the central regions of the galaxy only. Silva \& Wise
(1996) investigated the effects of centrally concentrated dust
distributions on the photometry of elliptical galaxies. They found
that, for models where the stars and dust have the same spatial
distribution or where the dust is more concentrated than the stars,
steep color gradients would be implied in the core, even for small
optical depths. However, Crane {\em et al.} (1993) and Carollo {\em et
al.} (1997) imaged the cores of a set of nearby elliptical galaxies
using HST, and both of them found relatively small color gradients,
and hence no direct sign for the presence of centrally concentrated
diffuse dust distributions. 


\begin{table}
\centering
\caption{Same as table~{\ref{param.tbl}}, but now for a fixed optical
depth $\tau=2$ and as a function of the dust exponent $\alpha$. The
last row, which is labeled ND, shows the same quantities for the
dust-free Plummer models.}
\begin{tabular}{cccccc}
\label{paramalpha.tbl}
$\alpha$ & $c_{\text{obs}}$ & $L_{\text{obs}}$ & $A$ & $\ell_{p,0}$ & $\sigma_{p,0}$\\
& (kpc) & (10$^9$\,L$_\odot$) & (mag) & (L$_\odot$/pc$^2$) & (km/s) \\ \\
1.0 & 5.00 &  4.60 & 1.09 &  58.6 & 178.0 \\
1.5 & 5.58 &  5.77 & 0.84 &  59.7 & 177.6 \\
2.0 & 5.91 &  6.75 & 0.67 &  61.5 & 177.1 \\
3.0 & 6.35 &  8.11 & 0.47 &  64.6 & 176.5 \\
5.0 & 6.69 &  9.55 & 0.29 &  68.8 & 176.1 \\[1mm]
ND  & 5.00 & 12.50 & 0.00 & 159.2 & 178.0
\end{tabular}
\end{table}


\begin{figure*}
\centering 
\includegraphics[clip,bb=40 488 554 685]{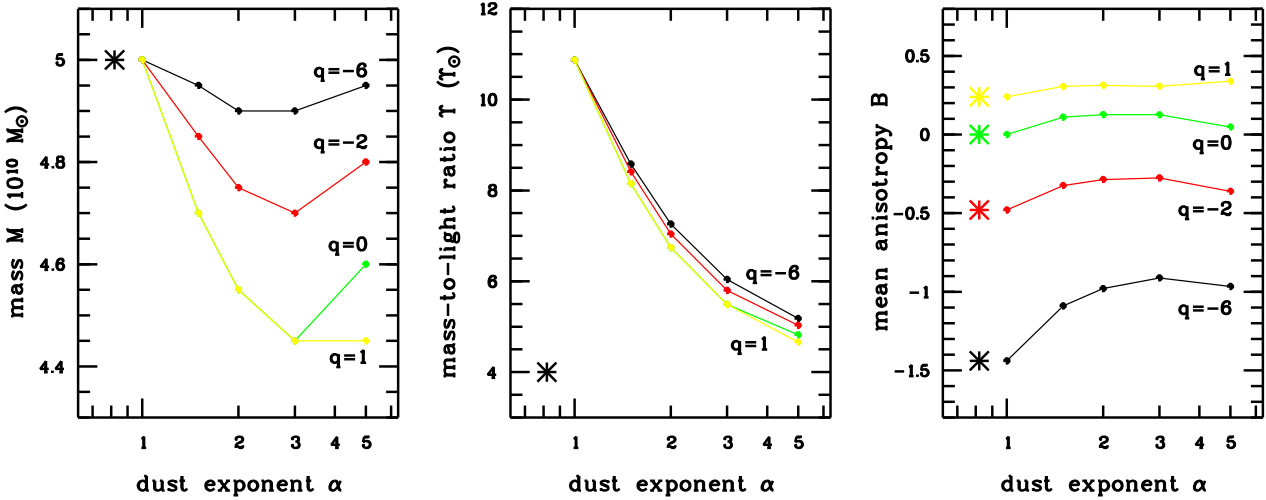}
\caption{The apparent mass $M$, the apparent mass-to-light ratio 
$\Upsilon$ and the apparent mean anisotropy ${\mathcal{B}}$, as a
function of the dust exponent $\alpha$, with $\tau=2$ fixed. They are
shown for the different orbital structures, ranging from very
tangential (black) to radial (light grey). The dots represent the
calculated values, the solid lines are just lines to guide the eye, in
contrary with figure~{\ref{deptau.ps}}. The asterisks represent the
intrinsic values of the represented quantities.}
\label{depalp.ps}
\end{figure*}


We consider $\alpha=1.5$, $\alpha=2$ and $\alpha=5$ (besides
$\alpha=3$) and create new dusty Plummer models for each of these
exponents and for the four orbital modes, where we fix the optical
depth at the median value $\tau=2$. Data sets are created and these
are modelled exactly as before. In particular, the observed light
profile can still in a satisfying way be approximated by a Plummer
potential~(\ref{Plpot}), for all values of $\alpha$ under
consideration. The observed core radii $c_{\text{obs}}$, as well as
some other parameters of the input models, are listed in
table~{\ref{paramalpha.tbl}} as a function of the dust exponent. In
figure~{\ref{depalp.ps}} we plot the results of our modelling: the
three plots show the apparent dynamical mass $M$, the apparent
mass-to-light ratio $\Upsilon$ and the apparent mean anisotropy
${\mathcal{B}}$ as a function of the dust exponent $\alpha$, for the
four different values of $q$. The intrinsic values of these quantities
are indicated by asterisks, in contrast with figure~{\ref{deptau.ps}},
where these correspond to the $\tau=0$ case.

Qualitatively, the effects are the same, independent of the value of
the dust exponent: the apparent dynamical mass decreases, the orbital
structure seems more radial and the apparent mass-to-light ratio
increases. Moreover, the dependence on the orbital structure of these
effect is independent of the dust exponent $\alpha$: the radial and
isotropic models tend to lose more mass and keep their orbital
structure, whereas for the tangential and very tangential models it is
vice versa.

Quantitatively, there is a dependence on the dust geometry. The
apparent dynamical mass and the apparent orbital structure are
unaffected for both very large or very small values of $\alpha$. The
effects appear to be strongest for $\alpha$ around 3. The
mass-to-light ratio however, has totally different dependence on
$\alpha$, it is strongly affected (a factor~2 to~3) for extended dust
distributions, and the effects decreases gradually when the dust
becomes centrally concentrated. In fact, $\Upsilon$ is largely
determined by the apparent luminosity, which is a strong function of
the dust geometry: the extinction is far more effective for extended
than for condensed distributions (paper~I, WS96).

This behavior can be illustrated when we consider the limits of very
low and very high dust exponents. For an extremely extended dust
distribution ($\alpha\rightarrow1$) on the one hand, the dust
effectively forms an absorbing screen with optical depth
$\tfrac{\tau}{2}$ between the galaxy and the observer. Relative to the
dust-free case, the light profile then decreases with a factor
$\exp\left(\tfrac{\tau}{2}\right)$, independent of the LOS, whereas
the morphology of $\ell_p(x)$ and the shape of the LOSVDs are not
affected (paper~I). Hence, a model with exactly the same potential,
dynamical mass and orbital structure, but with a mass-to-light ratio
which is a factor $\exp\left(\tfrac{\tau}{2}\right)$ higher, will fit
these data exactly. For an extremely centrally concentrated dust
distribution ($\alpha\gg5$) on the other hand, the dust effects will
be visible only in the very innermost regions. As a result, neither
the projected light density, nor the LOSVDs will be severely affected,
and hence the data will be nearly the same as in the dust-free
model. Hence all intrinsic quantities are recovered.

The question we want to answer in this chapter was whether the results
we obtained using the $\alpha=3$ model change dramatically in function
of the dust geometry. As we argued, neither centrally concentrated nor
very extended dust distributions seem very probable. Therefore we
consider the range $2\la\alpha\la5$, where the stars are somewhat more
concentrated than the dust, as representative for realistic
geometries. Although $M$, ${\cal{B}}$ and $\Upsilon$ do vary with
$\alpha$ in this range, the effects have qualitatively the same
behavior (decreasing apparent dynamical mass, increasing apparent
mean anisotropy), and quantitatively the same order of
magnitude. Hence we can conclude that our results of the $\alpha=3$
case, as summarized in table~{\ref{deptau.tbl}} can be considered as
representative.


\section{Discussion}


In this paper we investigated which errors can be made by not taking
dust into account in dynamical modelling procedures. Therefore we
created a set of galaxy models consisting of a dust and a stellar
component. We calculated the projected kinematics, taking dust into
account, using the method outlined in paper~I. These data sets are
then modelled as if no dust were present, and the {\em apparent}
dynamical properties of these models are calculated and compared to
the {\em intrinsic} ones, as a function of the orbital structure of
the input model, the optical depth of the dust and the dust geometry.

We find that (1) the dynamical mass of the galaxy tends to become
smaller, (2) the orbital structure seems to be radialized.  For a
fixed optical depth and dust geometry, the relative strength of these
effects depends on the orbital structure of the input model. For
radial and isotropic models the apparent mass decreases significantly,
with a typical amount of 5~per cent per optical depth unit, whereas
their orbital structure is hardly affected. For tangential models on
the other hand, the dynamical mass is less sensitive to the presence
of dust (about 3 per cent per optical depth unit for the tangential
model and less for the very tangential model), whereas now the
radialization is considerably stronger. Both effects are apparently
coupled, in a way that reminds of the mass-anisotropy degeneration in
spherical systems.

The effects are dependent on the shape of the dust distribution, but
not very critically: for the dust exponents in the range
$2\la\alpha\la5$ the effects are very comparable. On the one hand this
is fortunate, since it means that our calculations are more or less
model-independent, and can be applicated for a wide range of dust
geometries. On the other hand however, this means that dynamical
analyses will not be able to discriminate convincingly between
different values for the dust exponent. Analyses as ours can thus
hardly be used to constrain the distribution (and the origin) of the
smooth dust component in ellipticals. Further infrared and
submillimeter observations, in particular ISO data, are necessary to
solve this problem. Preliminary results include the detection of an
extended, very cold dust component in the dwarf elliptical NGC\,205
(Haas 1998), and of warm dust in the central regions of the Seyfert\,I
S0 galaxy NGC\,3998 (Knapp {\em et al.} 1996). A larger database of
ISO imaging of early-type galaxies at both mid-infrared an
far-infrared wavelengths would improve our knowledge significantly.

The combined results of paper~I and this paper may at first glance
seem quite contradictory. In paper~I we found that the observed
kinematics of elliptical galaxies are not severely affected by dust
obscuration. Whence it seemed obvious that modest amounts of dust do
not imply large uncertainties on dynamical mass determinations or
estimates of the anisotropy of these systems. In this detailed study
however we find that dust does have an important effect on the
determination of the dynamical structure, in particular the dynamical
mass and the anisotropy. The answer to this apparent discrepancy lies
in the fact that the potential plays am important role in the
determination of the internal structure of galaxies. Dejonghe \&
Merritt (1992) show that, in case of a spherical two-integral system,
the knowledge of the potential and the entire set of LOSVDs suffice to
determine the DF uniquely. The potential itself can be considerably
constrained by the LOSVDs, but is not uniquely determined, such that a
set of potentials will usually yield acceptable models for a data
set. Often one chooses that potential that is derived from the
observed light profile, if this one is one of the possible choices
(e.g.\ if dark matter is not assumed to play a major role). But if
dust is present, the light profile will be severely affected, even by
small amounts of dust, such that the matching potential will not be
the correct one. And as diffuse dust is assumed to be present in a
major fraction of the early-type galaxies, we argue that it is
important to at least be aware of its effects, which may not be as
trivial as one might imagine. As WS96 stressed, all broadband
observations of elliptical galaxies may be affected by dust, and hence
dust should be seriously taken into account in their
interpretation. We now can add that dust does also play a role in
dynamical analyses, and hence that it should also here be taken into
account, in a non-trivial way.

We close by giving a simple example to illustrate this point. A simple
way to estimate the mass of a gravitating system is its virial
mass. For example, Tonry \& Davis (1981) estimate the masses for a set
of 373 elliptical galaxies using a relation where mass is proportional
to the effective radius and the square of the central
dispersion. Although it is nowadays possible to obtain much better
mass estimates for nearby galaxies, the virial mass estimate is still
one of the only tools to constrain the mass of galaxies and clusters
at intermediate or high redshifts (Carlberg {\em et al.} 1996,
Carlberg, Yee \& Ellingson 1997, Tran {\em et al.}  1999). The
question is now how to correct these mass estimates for the presence
of dust\footnote{There is still no clarity about the amount and the
sources of dust grains in the intergalactic medium of galaxy
clusters. Studies investigating the extinction effects of background
galaxies and quasars yield controversial results (Ferguson 1993, Maoz
1995). Also the FIR emission is still inconclusive: recently, extended
ISO emission has been interpreted as evidence for the presence of
intracluster dust in the Coma cluster (Stickel {\em et al.} 1998) and
Abell 2670 (Hansen {\em et al.}  1999), but this evidence is still
controversial (Quillen {\em et al.}1999). The presence of
intergalactic dust still isn't firmly established, and, as Popescu
{\em et al.} (2000) suggest, the new generation of submillimeter
interferometers might contribute significantly to solve this
problem.}. A straightforward way is to estimate amount of dust using
IRAS or ISO data, and calculate the effects on dispersion and scale
length. Dispersions are only slightly affected by dust absorption
(paper~I), whereas scale lengths as the effective radius or the core
radius can increase substantially, as dust primarily removes light
from the centre of the system. Hence we find that the apparent mass of
the galaxy would increase as a function of $\tau$, while we find,
using detailed kinematic modelling, that the apparent mass decreases
with increasing optical depth. Moreover, this correction will be
independent of the orbital structure of the model. This again
illustrates the fact that dust effects are non-trivial and should be
fully taken into account.


\bsp


\begin{thebibliography}{}
\bibitem{BD99} Baes M., Dejonghe H., 2000, MNRAS, 313, 153 [paper~I]
\bibitem{BHR92} Bregman J.\,N., Hogg D.\,E., Roberts M.\,S., 1992, ApJ, 387, 484
\bibitem{BMC88} Bruzual G.\,A., Magris G.\,C., Calvet N., 1988, ApJ, 333, 673
\bibitem{CYEAGMP96} Carlberg R.\,G., Yee H.\,K.\,C., Ellingson E., Abraham R., Gravel P., Morris S., Pritchet C.\,J., 1996, ApJ, 462, 32
\bibitem{CYE97} Carlberg R.\,G., Yee H.\,K.\,C., Ellingson E., 1997, ApJ, 478, 462
\bibitem{CFIF97} Carollo C.\,M., Franx M., Illingworth G.\,D., Forbes D.\,A., 1997, ApJ, 481, 710
\bibitem{CSK...93} Crane P., Stiavelli M., King I.\,R., Deharveng J.\,M., Albrecht R., Barbieri C., Blades J.\,C., Boksenberg A., Disney M.\,J., Jakobsen P., Kamperman T.\,M., Machetto F., Mackay C.\,D., Paresce F., Weigelt G., Baxter D., Greenfield P., Jedrzejewski R., Nota A., Sparks W.\,B., 1993, AJ, 106, 1371
\bibitem{dJNHJ90} de Jong T., N{\o}rgaard-Nielsen H.\,U., Hansen L., J{\o}rgensen H.\,E., 1990, A\&A, 232, 317
\bibitem{D86} Dejonghe H., 1986, Phys.\ Rep., 133, 225
\bibitem{D87} Dejonghe H., 1987, MNRAS, 224, 13
\bibitem{D89} Dejonghe H., 1989, ApJ, 343, 113
\bibitem{DM92} Dejonghe H., Merritt D., 1992, ApJ, 391, 531
\bibitem{DD98} De Rijcke S., Dejonghe H., 1998, MNRAS, 298, 677
\bibitem{DDP89} Disney M.\,J., Davies J.\,I., Phillips S., 1989, MNRAS, 239, 939
\bibitem{EB85} Ebneter K., Balick B., 1985, AJ, 90, 183
\bibitem{FNC91} Fabian A.\,C., Nulsen P.\,E.\,J., Canizares C.\,R., 1991, A\&AR, 2, 191
\bibitem{F93} Ferguson H.\,C., 1993, MNRAS, 263, 343
\bibitem{FH91} Fich M., Hodge P., 1991, ApJL, 374, L17
\bibitem{FH93} Fich M., Hodge P., 1993, ApJ, 415, 75
\bibitem{F91} Forbes D.\,A., 1991, MNRAS, 249, 779
\bibitem{G93} Gerhard O.\,E., 1993, MNRAS, 265, 213
\bibitem{GdJ95} Goudfrooij P., de Jong T., 1995, A\&A, 298, 784 [GdJ95]
\bibitem{H98} Haas M., 1998, A\&A, 337, L1
\bibitem{HJN95} Hansen L., J{\o}rgensen H.\,E., N{\o}rgaard-Nielsen H.\,U., 1995, A\&A, 297, 13
\bibitem{HJNPGL99} Hansen L., J{\o}rgensen H.\,E., N{\o}rgaard-Nielsen H.\,U., Pedersen K., Goudfrooij P., Linden-V{\o}rnle M.\,J.\,D., 1999, A\&A, 349, 406
\bibitem{H75} Holmberg E., 1975, Stars and Stellar Systems IX, University of Chicago, 123
\bibitem{KGKJ89} Knapp G.\,R., Guhathakurta P., Kim D.-W., Jura M., 1989, ApJS, 70, 329
\bibitem{KGW92} Knapp G.\,R., Gunn J.\,E., Wynn-Williams C.\,G., 1992, ApJ, 399, 76
\bibitem{KRFHW96} Knapp G.\,R., Rupen M.\,P., Fich M., Harper D.\,A., Wynn-Williams C.\,G., 1996, A\&A, 315, L75
\bibitem{KX92} Kwan J., Xie S., 1992, ApJ, 398, 105
\bibitem{M95} Maoz D., 1995, ApJ, 455, L115
\bibitem{M98} Merluzzi P., 1998, A\&A, 338, 807
\bibitem{MM86} M\"ollenhof C., Marenbach G., 1986, A\&A, 154, 219
\bibitem{PTFV00} Popescu C.\,C., Tuffs R.\,J., Fischera J., V\"olk H., 2000, astro-ph/0001053, accepted for publication in A\&A
\bibitem{QRRCE99} Quillen A.\,C., Rieke G.\,H., Rieke M.\,J., Caldwell N., Engelbracht C.\,W., 1999, ApJ, 518, 632
\bibitem{RHBFJ91} Roberts M.\,S., Hogg D.\,E., Bregman J.\,N., Forman W.\,R., Jones C., 1991, ApJS, 75, 751
\bibitem{SW96} Silva D.\,R., Wise M.\,S., 1996, ApJ, 457, L71
\bibitem{SMG89} Sparks W.\,B., Macchetto F., Golombek D., 1989, ApJ, 217, 425
\bibitem{SLMHH98} Stickel M., Lemke D., Mattila K., Haikala L.\,K., Haas M., 1998, A\&A, 329, 55
\bibitem{TD81} Tonry J.\,L., Davis M., 1981, ApJ, 246, 666
\bibitem{TKvDFIM99} Tran K.-V.\,H., Kelson D.\,D., van Dokkum P., Franx M., Illingworth G.\,D., Magee D., 1999, ApJ, 522, 39
\bibitem{TM96} Tsai, J.\,C., Mathews W.\,G., 1996, ApJ, 468, 571
\bibitem{vdMF93} van der Marel R.\,P., Franx M., 1993, ApJ, 407, 525
\bibitem{vDF95} van Dokkum P.\,G., Franx M., 1995, AJ, 110, 2027
\bibitem{VV88} V\'{e}ron-Cetty M.-P., V\'{e}ron P., 1988, A\&A, 204, 28
\bibitem{WH95} Wiklind T., Henkel C., 1995, A\&A, 297, L71
\bibitem{WS96} Wise M.\,W., Silva D.\,R., 1996, ApJ, 461, 155 [WS96]
\bibitem{WS97} Wise M.\,W., Silva D.\,R., 1997, in The Nature of Elliptical Galaxies, Proceedings of the Second Stromlo Symposium, eds.\,M.\ Arnaboldi, G.\,S.\ Da Costa \& P.\ Saha, ASP Conference Series, vol.\ 116, p.\ 364  
\bibitem{WTC92} Witt A.\,N., Thronson H.\,A.\,Jr., Capuano J.\,M.\,Jr., 1992, ApJ, 393, 611
\end{thebibliography}
\end{document}